\documentclass[12pt,preprint]{aastex}

\usepackage{amssymb}
\usepackage{amsmath}
\newcommand{\be}{\begin{equation}}
\newcommand{\ee}{\end{equation}}

\shorttitle{Metal Free Gas at Redshift $z\approx6$} 
\shortauthors{Trenti et al.}

\begin{document}


\title{Metal-Free Gas Supply at the Edge of Reionization: \\ 
Late-Epoch Population III Star Formation }


\author{Michele Trenti} \affil{University of Colorado, Center for
  Astrophysics and Space Astronomy, 389-UCB, Boulder, CO 80309 USA}
\email{trenti@colorado.edu}
\and
\author{Massimo Stiavelli}
\affil{Space Telescope Science Institute, 3700 San Martin Drive, Baltimore, MD 21218 USA}
\email{mstiavel@stsci.edu} 
\and

\author{J.~Michael Shull} \affil{University of Colorado, CASA, Dept.
  of Astrophysical \& Planetary Sciences, 389-UCB, Boulder, CO 80309
  USA} \email{michael.shull@colorado.edu}


\begin{abstract}

  While the average metallicity of the intergalactic medium rises 
  above $Z \gtrsim 10^{-3} Z_{\sun}$ by the end of the reionization,
  pockets of metal-free gas can still exist at later times. We
  quantify the presence of a long tail in the formation rate of
  metal-free halos during late stages of reionization (redshift $z\approx 6$),
  which might offer the best window to detect Population III stars.
  Using cosmological simulations for the growth of dark matter halos, 
  coupled with analytical recipes for the metal enrichment of their
  interstellar medium, we show that pockets of metal-free gas exist
  at $z\approx 6$ even under the assumption of high efficiency in
  metal pollution via winds. A comoving metal-free halo formation rate 
  $d^2 n / dt\, dV \gtrsim 10^{-9}~ \mathrm{Mpc^{-3}~yr^{-1}}$ is
  expected at $z=6$ for halos with virial temperature $T_{vir} \approx
  10^4$ K (mass $\sim 10^8 ~\mathrm{M_{\sun}}$),
  sufficient to initiate cooling even with strong negative radiative
  feedback. These halos will appear as absorption systems with a
  typical hydrogen column density of $ \sim 10^{20}$ cm$^{-2}$, a
  sky covering factor $5 \times 10^{-3}$ and a number density of $25$
  arcmin$^{-2}$ for $5.5 \leq z \leq 6.5$. Under the assumption of a
  single Population III supernova formed per metal-free halo, we
  expect an observed supernova rate of $2.6\times
  10^{-3}~\mathrm{deg^{-2}~yr^{-1}}$ in the same redshift range. These
  metal-free stars and their supernovae will be isolated and outside
  galaxies (at distances $\gtrsim 150~h^{-1}$~kpc) and thus
  significantly less biased than the general population of $\sim
  10^{8}~ \mathrm{M_{\sun}}$ halos at $z \approx 6$. Supernova searches
  for metal-free explosions must thus rely on large area surveys. If
  metal-free stars produce very luminous supernovae, like SN2006gy,
  then a multi-epoch survey reaching $m_{\rm AB} =27$ at 1 $\mu$m is sufficient
  for detecting them at $z=6$. While the \emph{Large Synoptic Survey 
  Telescope} (LSST) will not reach this depth in the z band, it will be
  able to detect several tens of Population III supernovae in the i
  and r bands at $z \lesssim 5.5$, when their observed rate is down to
  $3-8\times 10^{-4}~\mathrm{deg^{-2}~yr^{-1}}$.

\end{abstract}


\section{Introduction}

Metal-free stars, conventionally called Population III stars,
represent the first step to exit the dark ages of the universe and
start the reionization era. The chemical enrichment of the
intergalactic medium (IGM) opens the way to the formation of
stars and galaxies like our own Milky Way (see \citealt{bromm04}). 
This first step in the star formation history of
the Universe has attracted large theoretical interest, and thanks to
a combination of analytic and numerical investigations, we now have a
reasonably solid understanding of the properties of Population III
stars within the $\Lambda CDM$ paradigm. Population III stars are
expected to start forming within dark matter halos of mass $\sim 10^5~
\mathrm{M_{\sun}}$ (minihalos) at $z \approx 60$, about $40~
\mathrm{Myr}$ after the Big-Bang \citep{naoz06,ts07a}, most likely
with a top-heavy mass function and just one star per halo
\citep{abel02,bromm04,oshea07,oshea08}. Once the epoch of Population
III stars begins, their formation rate in minihalos increases
exponentially with redshift, reaching a level of $10^{-7}~
\mathrm{Mpc^{-3}~ yr^{-1}}$ at $z \approx 25$. Eventually, by $z
\approx 15$, Population III star formation in minihalos is 
suppressed by the build-up of a strong radiative background in the
H$_2$ Lyman-Werner (LW) bands, which photodissociates the molecular
hydrogen responsible for the cooling of the primordial gas. A second
mode of metal-free star formation is possible in larger halos ($M
\gtrsim 2 \times 10^{7}~ \mathrm{M_{\sun}}$) where self-shielding
against LW feedback is more effective and where cooling may also be
driven by Ly$\alpha$ lines. However as the redshift decreases, it
becomes progressively more unlikely that $\sim10^{7}~
\mathrm{M_{\sun}}$ halos composed of metal-free gas can form
\citep{furlanetto05,ts09}. The transition to Population II star
formation happens in the metallicity range 
$Z \sim 10^{-3}-10^{-4}~Z_{\sun}$ \citep{bromm01a,santoro06,smith_b09} 
or even at lower metallicity 
in the presence of dust \citep{schneider06}. Pockets of pristine gas are
still likely to exist at $z<6$, thanks to the inhomogeneity of
metal enrichment in the IGM
(\citealt{scannapieco03,schneider06,tornatore07};
\citealt{ricotti08}).

An unambiguous observational test of the Population III paradigm would
represent a significant step toward our understanding of star
formation in the Universe. This is still missing, despite many
attempts ranging from Galactic surveys for metal-poor stars
\citep{beers05,frebel07,cohen08} to surveys aimed at identifying their
spectral signatures in galaxies at very high redshift
\citep{nagao05,nagao08}. In fact, Population III stars are expected to
be extremely hot and thus strong emitters in the He~II lines
\citep{tumlinson00,oh01}. Some indirect evidence has also been
proposed. For example, \citet{jimenez06} argue that many puzzles
related to Lyman Break Galaxies at $z \gtrsim 4$ would be solved by
assuming that a fraction of their light is produced by metal-free
stars. Similarly, \citet{malhotra02} suggest that the large equivalent
width of Ly$\alpha$ emitters at $z>5$ might be a signature of
Population III stars (\citealt{tumlinson03,scannapieco03}). However
co-existence of both Population III and II stars in the same galaxy
appears to require a fine tuning of the star formation history,
especially because the initial mass function (IMF) of metal-free stars
is probably more biased toward large masses than that of Population II
stars. {In addition, it would be extremely hard to avoid mixing of 
  of metal-free and metal polluted gas in a galaxy, because of the
  turbolent nature of the ISM (e.g. see
  \citealt{balsara05,balsara08}). Therefore a significant population
  of primordial stars in Lyman Break Galaxies is unlikely. } Another
indirect fingerprint of Population III stars is the presence of
peculiar abundance patterns in the gas chemically enriched by
Population III stars, with enhanced even-odd element abundance ratios
\citep{heger02,tumlinson04}, but again no strong evidence is
available. Finally, the transition from Population III to Population
II star formation leaves an imprint in the reionization history of the
Universe (\citealt{shull08}; see also \citealt{bagla09}).

The null result achieved so far is not surprising. In fact, the task
is very challenging: the expected top-heavy IMF implies that the
lifetime of a single Population III star is rather short --- possibly
only a few million years --- and they form in isolation, at least in
minihalos \citep[e.g., ][]{oshea08}. These objects are extremely faint
at cosmological distances ($m_{\rm AB}>40$ at $z>10$) and thus
unlikely to be detected directly via imaging even with future
facilities such as the \emph{James Webb Space Telescope} (JWST). A
better chance of detection might be afforded by pair-instability
supernovae. In fact, metal-free stars in the mass range $140-260~
\mathrm{M_{\sun}}$ end their lives with powerful explosions (up to
$10^{53}$ ergs released) powered by the production of $e^+ e^-$ pairs
\citep{heger02}. These are up to 10 times more luminous than standard
SNs from Population III stars of lower mass \citep{heger02}, although
also metal-free stars with $\sim 30-100~ \mathrm{M_{\sun}}$ might
produce luminous hypernovae with an energy release above $10^{52}$
ergs \citep{nomoto03}. Still, many observational challenges are
present at such high redshift ($z \gtrsim 10$), especially because of
the significant time dilation of their light curves (e.g. see
\citealt{scannapieco05}; see also \citealt{mesinger06}).

In this paper, we aim to evaluate the number density of pockets of
primordial gas that survive to $z<10$, using conservative assumptions
on the metal enrichment efficiency and parameterizing the results
based on the strength of galactic winds. Even with efficient metal
transport, metal-free pockets still survive at $z \approx 6$. Thus if
massive stars are formed out of this primordial gas,
the expected star formation rate can lead to the detection of
hypernovae using all-sky surveys such as the {\it Large Synoptic Survey
Telescope} (LSST).

The paper is organized as follows. 
In \S ~\ref{sec:sim} we present our model for the formation rate
of metal-free halos around the end of the reionization era, whose
predictions are discussed in \S ~\ref{sec:predictions} along with
the prospects for the observational measures of the metal-free star
formation rate after reionization. We conclude in \S ~\ref{sec:conc}.

\section{Pristine halo formation rate}\label{sec:sim}

\subsection{Numerical setup}\label{sec:model}

The numerical simulations have been carried out using the Particle
Mesh Tree code {\it Gadget-2} \citep{springel05}. We adopt a cosmology 
based on the fifth year WMAP data \citep{komatsu08}:
$\Omega_{\Lambda}=0.72$, $\Omega_{m}=0.28$, $\Omega_{b}=0.0462$,
$H_0=70~$ km~s$^{-1}$~Mpc$^{-1}$ and $\sigma_8=0.817$, where
$\Omega_m$ is the total matter density and $\Omega_b$ is the baryon
matter density in units of the critical density, $\rho_{c}= 3H_0^2/(8
\pi G)$. Here, $H_0$ is the Hubble constant parameterized as $H_0 =
100~ h$~km~s$^{-1}$~Mpc$^{-1}$, $\Omega_{\Lambda}$ is the dark energy 
density, and $\sigma_8$ is the root mean squared mass fluctuation in a 
sphere of radius $8~h^{-1}$~Mpc extrapolated to $z=0$ using linear theory.  
The initial conditions have been generated with a code based on the Grafic
algorithm \citep{bertschinger01} using a $\Lambda CDM$ transfer function
computed via the fit by \citet{eisenstein99} with spectral index $n_s=0.96$.

We start our principal simulation at redshift $z=199$ using a box of
edge $l=7~h^{-1}$~Mpc, $N=1024^3$ particles, a mass
resolution of $3.4 \times 10^{4}~ \mathrm{M_{\sun}}$ and a force
resolution of $0.16~h^{-1}$~kpc. A total of 79 snapshots are
saved from $z=30$ to $z=4$, and from these snapshots dark matter halos
are identified with a friend-of-friend halo finder \citep{davis85}
using a linking length equal to 0.2 of the mean particle
separation. An additional, lower resolution simulation ($N=512^3$,
$l=10~h^{-1}$~Mpc) is described in Appendix ~\ref{sec:validation}, 
where we discuss the convergence of our results.

For every redshift $z$, we define a minimum halo mass for primordial
star formation which is the lowest halo mass that can efficiently cool
in a Hubble time. Two main coolants are available for metal-free gas:
molecular hydrogen and atomic hydrogen \citep[see][]{tegmark97} with
different efficiencies depending on the gas temperature. At the time
of formation of the halo the gas temperature is approximately equal to
the halo virial temperature, 
\be \label{eq:tvir}
T_{\rm vir}(M,z) = 2554~ \mathrm{K} \left(\frac{M}{10^6~ \mathrm{M_{\sun}}}\right)^{2/3}
\left(\frac{1+z}{31}\right).
\ee

\begin{figure} 
  \plotone{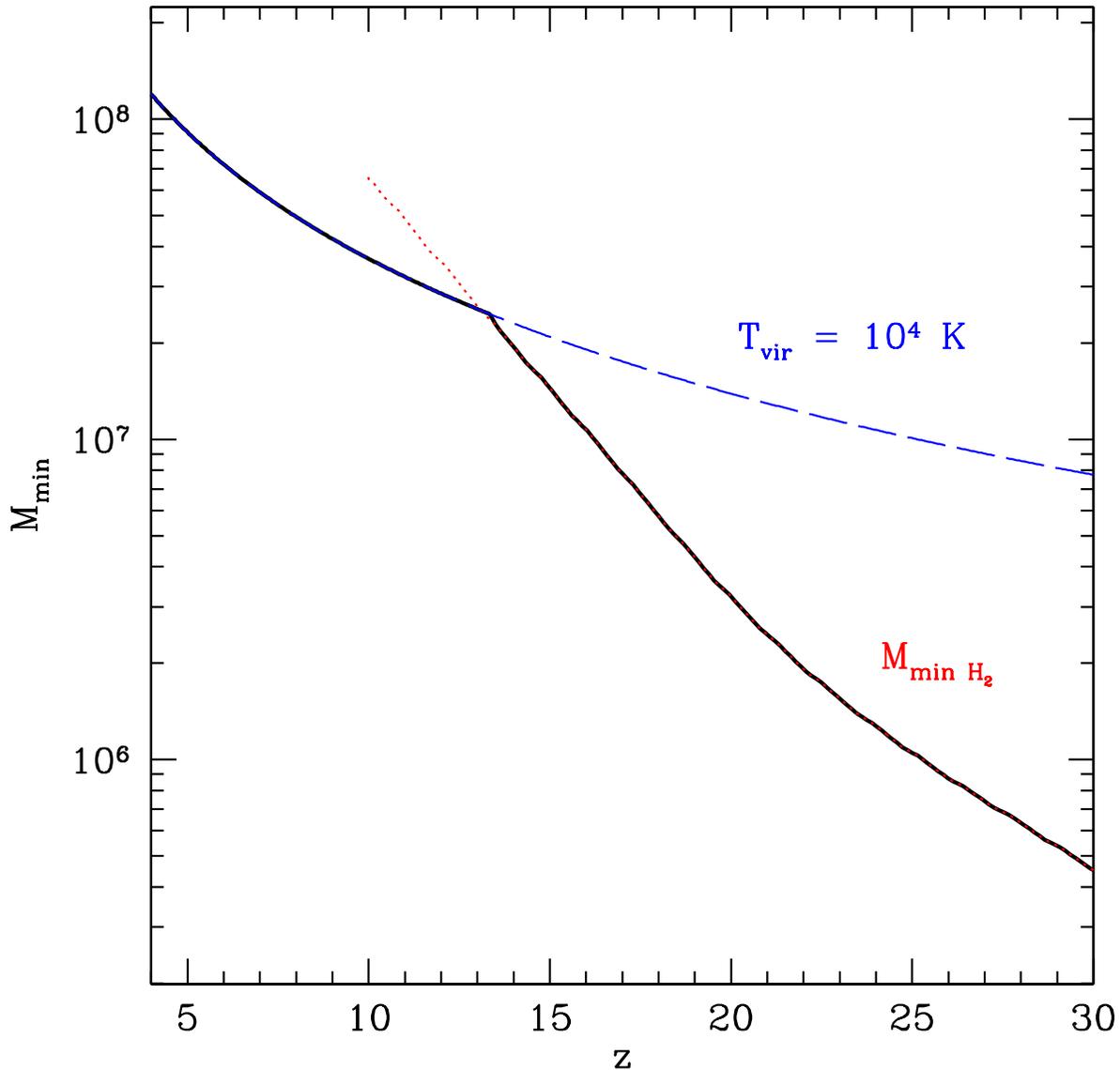} \caption{Black solid line: minimum halo mass
  required for cooling and collapse of a metal-free halo in presence
  of an H$_2$ dissociating Lyman-Werner background. Calculations are 
  based on the standard model by \citet{ts09} --- see their Fig.~1. 
  For $z<13.4$, molecular hydrogen cooling is strongly suppressed in 
  minihalos, and Population III star formation occurs in halos with virial
  temperature $T_{\rm vir} =10^4 K$ (dashed blue line). At higher redshift,
  cooling is instead possible in minihalos with a minimum mass $M_{min
  H_2}$ (red dotted line) self-consistently derived from the 
  Lyman-Werner background created by the first stars and galaxies (see
  \citealt{ts09} for details of the model).}\label{fig:minmass}
\end{figure}

Molecular hydrogen is an effective coolant at virial temperatures
$T_{\rm vir} \gtrsim 10^3$ K \citep{galli98}, but it can be
photodissociated by radiation in the H$_2$ Lyman-Werner bands
\citep{lepp83}, which progressively builds up as the redshift
decreases and the Universe starts to be ionized. Extended
Press-Schechter modeling shows that this cooling channel is
significantly quenched at $z \gtrsim 15$ \citep{haiman97,haiman06,ts09}, at which point
metal-free star formation shifts to halos with $T_{\rm vir}\gtrsim 10^4$ K, 
where self-shielding from the Lyman-Werner background is efficient and 
atomic hydrogen cooling is also possible. We therefore assume that the 
minimum halo mass capable of cooling is that derived in the
reference model of \citet{ts09}, which includes the effect of a
self-consistent Lyman-Werner background. This minimum mass versus
redshift, $M_{\rm min}(z)$, is presented in Fig.~\ref{fig:minmass}. 
Given the mass
resolution of our simulation, we can resolve halos capable of forming
Population III stars via H$_2$ cooling up to $z \leq 26$ with
$N \geq 32$ particles. For $z \leq 13$ the minimum halo mass capable
of forming stars corresponds to that of halos with $T_{\rm vir} \geq 10^4~
\mathrm{K}$, which can be written as:
\be
M_{\rm min}(z \leq 13) = \left ( 7.75 \times 10^6~\mathrm{M_{\sun}} \right) 
  \left(\frac{1+z}{31}\right)^{-3/2} .
\ee

As a conservative upper limit to the self-enrichment of halos, we flag
all particles of a halo with mass $M>M_{\rm min}(z)$ as ``metal
enriched''. We further assume that a necessary, but not sufficient
condition for a halo to be pristine is to be made entirely of
particles that have never been in a halo capable of cooling at higher
redshift.

Metal enrichment is also possible due to supernovae winds originating
in nearby halos. Wind bubbles during the reionization epoch can
propagate out to a comoving radius of $\sim 150~h^{-1}$~kpc by
redshift $z=6$ accordingly to the \citet{furlanetto03} model (but
\citealt{madau01} predict bubble radii approximately two times
smaller). An estimate of the wind bubble radius can be also
obtained noting that a bubble expanding at $\sim 60~ \mathrm{km~s^{-1}}$
\citep{furlanetto03} reaches a radius of $\sim 150~h^{-1}$~kpc 
after 500~Myr. Metals are in fact assumed to be transported
by winds at high speed ($\sim 300~\mathrm{km~s^{-1}}$) until they mix
in a thin spherically symmetric shell, propagating at $\sim 60$
km~s$^{-1}$, that eventually fragments by cooling and pollutes the IGM
\citep{furlanetto03}. The simple picture discussed here is consistent
with more detailed modeling of the interaction of the winds with the
IGM. Using the non-cosmological Sedov-Taylor model, as written in
equation 8 of \citet{tumlinson04}, we obtain bubble sizes of $\lesssim
150~h^{-1}$~kpc at $z\sim 6$.

To derive a conservative upper limit to the wind enrichment, we
compute for every halo that is not self-enriched the distance to
neighbor halos that have experienced star formation in the past, that
is of mass $M>M_{\rm min}$. The size of the wind bubble around these halos
depends on the first time they experienced star formation, assuming a
bubble expansion speed in the range $v_{\rm wind} = 40-100$
km~s$^{-1}$. If the distance between the non-self-enriched halo and
other halos is greater than the wind bubble sizes, we have a truly
pristine halo. No metal wind had the opportunity to reach the halo
(see also the discussion of metal enrichment in \citealt{ricotti08}).

This assumption is in reality likely to overestimate the efficiency of
wind pollution. As the wind bubble expands, its capability of
enriching the IGM above the critical metallicity needed to quench
Population III formation is progressively reduced. For example, if we
assume a critical metallicity of $10^{-3.5}~ Z_{\sun}$, a non
self-enriched halo of mass $10^8~ \mathrm{M_{\sun}}$ needs to be
polluted with $\sim 100~ \mathrm{M_{\sun}}$ of metals by winds. 
We assume that the galaxy with the winds has a star formation efficiency
$\epsilon = 10^{-2}$ and a metal yield of $Y \sim 2.4 \times 10^{-2}$
(total mass of metals released per unit solar mass in the IMF).  This
value is estimated by assuming a $3~ \mathrm{M_{\sun}}$ metal release 
from a supernova and one supernova formed for every $120~ \mathrm{M_{\sun}}$ 
of mass in the IMF.  The wind then contains about $3960~
\mathrm{M_{\sun}}$ of metals for a wind galaxy also hosted in a $10^8~
\mathrm{M_{\sun}}$ halo. Therefore a non self-enriched halo needs to
have a cross section with the wind bubble of at least $100/3960 \times
4 \pi \approx 0.32$ sr in order to be polluted above the critical
metallicity assuming a 100\% efficiency in the pollution
process. Given that a $10^8~ \mathrm{M_{\sun}}$ halo has a (comoving)
virial radius $\sim 10.5~h^{-1}$~kpc at $z=6$, the maximum radius of
the wind bubble is only $ \sim 33~h^{-1}$~kpc. Of course, multiple
bubbles can contribute to the metal enrichment of a halo, and this
enrichment can happen before the virialization process is completed.
The gas is then less concentrated and has a larger cross section.


\subsection{Results}

In Fig.~\ref{fig:genetic_pristine} we plot the formation rate 
($d^2 n /dt \, dV$) of halos that are not self-enriched. This quantity is
slowly declining from $z=25$ to $z=13$ and exhibits a local maximum at
$z \sim 10$, at which time about 30 halos are formed per year per
comoving Gpc$^3$. Interestingly, their formation rate is only slowly
varying in the redshift range $7 \lesssim z \lesssim 25$ because of
two competing effects. First, overdensities that lead to these halos
become progressively more efficient, and second, as time passes, the
supply of metal-free gas decreases and it is more probable that a
region has been self-enriched. The halo formation rate starts dropping
rapidly only at $z \lesssim 7$. The evolution of $d^2 n /dt \, dV$
measured in our numerical simulation is in excellent agreement with
the \citet{ts09} model, shown as the red-dashed line in
Fig.~\ref{fig:genetic_pristine}. There is a discrepancy by a factor
$\sim 2$ at $z \gtrsim 15$, probably related to the finite
resolution of the numerical simulation which does not allow us to
resolve the progenitor of halos formed at these redshifts. The
agreement improves to $\sim 30$\% at lower redshift. With the aid of
the analytic model, we can understand the origin of the local maximum
in the halo formation rate at $z \sim 10$ (see Fig.~1 in
\citealt{ts09}) as associated with the peak of Population III stars
formed in halos with $T_{\rm vir}= 10^4$ K. The decline from $z\sim 25$ to
$z\sim 13$ is instead associated with Population III formation in
minihalos, progressively quenched by the rising Lyman-Werner
background.

The spatial distribution of these halos is consistent with that of the
general population of halos with the same mass, independent of their
enrichment history, as we have verified by measuring the two-point
correlation function and thus their bias. This result is natural in
the context of extended Press-Schechter modeling, because
self-enrichment depends only on density fluctuations at scales smaller
than the halo mass. 

Dependence on the environment is present instead in the case of wind
enrichment. To illustrate the effect of wind enrichment, we plot in
Fig.~\ref{fig:nearest_neighbors} a distance vs. mass scatter plot for
the nearest neighbors of non self-enriched halos formed at
$z=6$. The halos closest to those that have not been
self-enriched are typically small ($10^8~ \mathrm{M_{\sun}}$) and at 
distances comparable to those over which
the wind bubbles can propagate at $z=6$. Metal enrichment of non-self
enriched halos is thus dominated by dwarf-like galaxies, similarly to
what happens in the local universe (\citealt{stocke06}; see also
\citealt{fujita04}). About 20\% of the non self-enriched halos
have no neighbors within $150~h^{-1}$~kpc, a distance that essentially
guarantees that these halos contain genuine metal-free gas. Depending
on details of the wind propagation, up to 50\% of these halos can
be intrinsically metal free. The median of the distance distribution
in Fig.~\ref{fig:nearest_neighbors} is $82.4~h^{-1}$~kpc, for which a
single wind bubble is unlikely to chemically enrich a
metal-free halo above the critical metallicity for quenching
Population III formation. To quantify the wind enrichment we
compute the wind bubble size for all the star-forming halos in
our simulation as described in \S~\ref{sec:model}. We plot the
resulting metal-free halo formation rate in
Fig.~\ref{fig:final_popIII_sfr} for three different values of the wind
bubble expansion velocity, $v_{\rm wind} = 40,60,100$ km~s$^{-1}$. Even in the
``worst case scenario" of propagation at 100 km~s$^{-1}$, metal-free halos
continue to exist down to $z \sim 5$.

\subsection{Absorption Systems Predictions}

These metal-free halos leave an observational signature as absorption 
systems before star formation is associated with them. To estimate their 
column density, we assume that the baryon profile follows the dark matter, 
a reasonable assumption immediately after virialization and before 
significant cooling takes place. Assuming a helium mass fraction 
$Y=0.2477$ and a WMAP-5 cosmology \citep{komatsu08}, we can write the 
typical hydrogen column density of an halo of total mass $M_h$ as: 
\begin{equation}
\langle N_H \rangle = 10^{20}~\mathrm{cm^{-2}} \left ( 
\frac{\Omega_m \mathrm{h^2}}{0.1372}\right) \left ( 
\frac{M_h}{10^8~ \mathrm{M_{\sun}}} \right)^{1/3} \left ( 
\frac{1+z}{7}\right)^2.
\end{equation}
If the gas spends 5 Myr before cooling and initiating the protostellar 
core formation, we estimate from Fig.~\ref{fig:final_popIII_sfr} that the 
number density of metal-free absorbers is $\sim 25$ arcmin$^{-2}$ at $z=6$. 
Their physical size is 
\begin{equation}
  R_{\rm vir} = \left[ \frac {3 M_h}{4 \pi \rho_{\rm vir}} 
      \right]^{1/3} \approx (15.3~{\rm kpc}) (1+z)^{-1} 
      \left( \frac{M_h}{10^8~ \mathrm{M_{\sun}}} \right)^{1/3} \; ,  
\end{equation} 
and their covering factor on the sky is $\sim 5 \times 10^{-3}$ for $z
\in [5.5:6.5]$. The column density of these system is at the edge of
the Damped Ly$\alpha$ (DLA) limit and thus well within the
observational limits of both Ly$\alpha$ absorption or 21-cm emission
studies. The main limitation for the detection in absorption is the
very low number of bright QSOs known at $z>6$ \citep{fan06}. In
addition, the number density of metal-enriched DLAs is higher by a
factor of $\sim 300$ compared to that of metal-free systems if we
extrapolate to $z \sim 6$ the intrinsic density of neutral gas found
at $z \sim 4.5$ in DLAs ($\Omega_{H I}^{(DLA)}(z=4.5) \sim 10^{-3}$;
see \citealt{trenti06}) Obtaining strong limits on the metallicity of
a $z \sim 6$ absorber will be challenging even for a 30 m class
telescope.

\section{Population III stars and prospects for their detectability}
\label{sec:predictions}

From the metal-free halo formation rate we can derive a Population III
star formation rate once we assume an efficiency for the conversion of
gas in stars, a poorly constrained parameter. Numerical
simulation of the collapse and cooling of metal-free halos seem to
favor massive Population III stars formed in isolation, even in halos
with virial temperature $T_{\rm vir} \approx 10^4$ K in the presence of a
strong Lyman-Werner background (see \citealt{oshea08}). However, these
simulations are limited to the formation of protostellar cores and are
not able to reach the point of actual creation of the star. In
addition, cosmological simulations focus on Population III star
formation at the start of reionization at $z\gtrsim 15$. The physical 
conditions at the end of the reionization epoch are different. In 
particular, the CMB temperature floor is significantly lower, and a 
strong ionizing UV background is present, which  
may lead to a change in the initial mass function of Population
III stars as suggested by \citet{yoshida07}. Late time metal-free
stars could have masses of 30-50 $M_{\sun}$, rather than the values 
$M \gtrsim 100 M_{\sun}$ suggested at higher redshift. 
Note also that, while the very massive star hypothesis is a
strong prediction of numerical simulations of Population III star
formation, there is some tension with the mass function needed to
explain the abundance patterns of extremely metal poor stars in the
Milky Way (see \citealt{tumlinson06}). Here we explore these two
different scenarios for the formation of metal-free stars at the edge
of the reionization, by assuming (i) very massive stars formed in
isolation and (ii) clusters of massive stars.

\subsection{Detection during the main sequence lifetime}

First, we evaluate the luminosity of Population III stars on the
main sequence. Under the assumption that these stars are massive, we 
find their main sequence luminosity to be $3 \times 10^{21}~
\mathrm{erg~s^{-1}~Hz^{-1}~M_{\sun}^{-1}}$ \citep{bromm2001}, which
translates into a 1 $\mu$m observed magnitude $m_{\rm AB} \approx 40$ 
at $z=6$ for a $100 ~M_{\sun}$ star. Therefore, direct detection will 
be extremely challenging (see Table~\ref{tab:obs}).
This is essentially hopeless if star formation happens in isolation.  
Even if the massive star formation formation occurs in clusters, 
significant problems are present. For example, a cluster formed with
star formation efficiency $\epsilon = 10^{-2}$ in a $10^8~
\mathrm{M_{\sun}}$ halo a would have an observed 1 $\mu$m luminosity 
$m_{\rm AB} \approx 32$.

In our model with $v_{\rm wind} = 60$ km~s$^{-1}$, we expect a number
density on the sky of $\sim 10$ arcmin$^{-2}$
i-dropouts\footnote{These are high-redshift sources identified in
  broad-band imaging by inferring the location of the Ly$\alpha$
  IGM-absorption break, see \citet{steidel96}.} Population III sources
at $5.5 \lesssim z \lesssim 6.5$ (assuming that each source remains
visible for 2 Myr). Detecting them requires significantly deeper
exposures than the HST-UDF, as those observations only reach down to
$z_{850LP} = 28.6$ at $5\sigma$ \citep{oesch09}. Even JWST might not
be sufficient, unless the Population III star formation efficiency is
higher than $\epsilon = 10^{-2}$ \citep[see also][]{johnson09}. In
fact, a deep-field JWST integration is expected to reach $m_z \approx
31$ in 65 hours (see Table~\ref{tab:obs}). Adopting a different
strategy and searching for Ly$\alpha$ emission via narrow-band filters
would require a sensitivity of $3 \times 10^{41}~ \mathrm{erg~
  s^{-1}}$ for the same sources \citep{bromm2001} with an expected
number density $\sim 0.5$ arcmin$^{-2}$. The lower number density on
the sky is given by the smaller redshift range probed by narrow band
filters ($\Delta z \sim 0.05$ versus $\Delta z \sim 1$ in broad-band
imaging). Current 8 m class telescopes reach a sensitivity of
$10^{43}~ \mathrm{erg~s^{-1}}$ at $z\sim 7$, equivalent to an observed
flux of $2 \times 10^{-17}$ erg s$^{-1}$ cm$^{-2}$ \citep{ota08}. A 30
m class telescope is expected to improve this limit by about a factor
50, reaching $\sim 10^{-19}$ erg s$^-1$ cm$^{-2}$, which might be
sufficient to detect Ly$\alpha$ emission from a large cluster of
Population III stars.

Even if future telescopes reach the sensitivity required to detect
Population III clusters, the main problem is how to distinguish them
from Population II sources. Adopting the luminosity function measured
at redshift $z=6$ from the UDF survey \citep{bouwens06} and
extrapolating it down to $m_{z} = 32$, we expect a number density of
$\sim 42 $ arcmin$^{-2}$ i-dropouts from metal-enriched galaxies.
While contamination from Population II stars can in principle be
addressed owing to the higher effective temperature of Population III
stars \citep{tumlinson00}, it will be extremely challenging to
identify the characteristic He~II emission features expected from
higher temperatures, for sources near the detection limit of an
imaging-only survey. Therefore, disentangling the relative
contribution does not appear feasible, especially given the intrinsic
uncertainties on the luminosity function at high redshift
\citep[see][]{ts07b}. Although Population III sources at lower
redshift ($z\approx5$) will appear brighter than their $z\approx 6$
counterparts, their direct identification as V-dropouts will be even
more challenging because of the increased number of Population II
V-dropout sources combined with a sharp decline of the number of
Population III sources at this redshift (see
Fig.~\ref{fig:final_popIII_sfr}).

\subsection{Supernovae Detection}

The possibility of detecting Population III stars in their Supernova
phase are more promising, especially if these stars are very massive
and hypernovae and/or pair instability SNe (PISNe) are preferentially
produced. Theoretical modeling of PISNe light curves by
\citet{scannapieco05} predicts a peak luminosity $M_{\rm AB} \in
[-19:- 21.5]$, depending on the mass of the progenitor ($M \in
[140:260] \mathrm{M_{\sun}} $). This is in reasonably good agreement
with the most luminous SN observed to date, SN2006gy
\citep{smith2007}, {but the results by \citet{heger02} and
  \citet{scannapieco05} are based on one-dimensional, non-rotating
  simulations. Thus they carry some uncertainty both in the
  expected mass range for PISN explosions as well as in the total
  energy released. For example, \citet{ekstrom08} find that inclusion
  of significant amounts of rotation could enhance the explosion
  energy.} Assuming {here} the observed light curve of SN2006gy as a
reference (but note that SN2006gy is not metal free), we expect a peak
luminosity $M_{\rm AB} \sim - 21.8$, which translates into a 0.14
$\mu$m $M_{\rm AB} \sim - 20.5$ assuming a blackbody spectrum with
effective temperature $T_{\rm eff} = 15000$ K \citep{scannapieco05}.
This leads to an observed 1 $\mu$m $m_{\rm AB} \approx 26.25$ at
$z\sim 6$ (distance modulus of $48.86$ and K-correction of $-2.5
\log_{10}{(1+z)}$). To obtain a robust detection of the supernova peak
luminosity curve, a survey reaching down to $m_{\rm AB} \approx 27.0$
might be adequate, although the near-UV spectrum of SN2006gy is not
well constrained. Thus the prediction of its observed luminosity is
intrinsically uncertain at $\lambda \lesssim 1 \mu m$ when the source
is at $z\sim 6$. Because a pair instability supernova light curve
remains near its peak for a relatively long time (two to three months)
plus cosmological time dilation for the observer rest frame, a search
for $z\sim 6$ SNs would require a multi-year time-frame. This is
convenient, as it allows stacking multiple images for recurring
all-sky surveys such as LSST. Given the current LSST requirements
\citep[see][]{ivezic08} a $5 \sigma$ detection at $m_{\rm AB} = 26.25$
in the z band cannot be reached within the main survey, but is within
reach of a deeper ``mini-survey'' (see \citealt{ivezic08} Sec. 3.2.2).

Supernovae detection at $z \sim 5$ does appear potentially within the
capabilities of the main LSST survey, even though these sources will
be rarer than their $z\sim 6$ counterparts. Population III supernovae
at $4.5 \lesssim z \lesssim 5.5$ can be detected with LSST in the
i-band, where the instrument is more sensitive by 0.7 magnitudes than
in the z-band. In addition, the $z\sim 5$ sources will be about 0.3
magnitudes brighter compared to those at $z \sim 6$. A $4 \sigma$
limit of $m_{i} = 25.95$ can be achieved by stacking the 23 individual
exposures, each with sensitivity of $24.0$, that are planned for each
year of operations (see Table~\ref{tab:obs}). Combining the i-band
data with those in the z-band will further increase the confidence
level at which a source is identified. This strategy will push the
observations to their detection limit, but if there is a distribution
of peak luminosity, LSST may be able to identify at least the
brightest explosions. Of course, with the very large-data data set
generated by LSST, some false positives are expected. Follow-up of the
candidates will screen out random fluctuations in the photometry.
Identifying these transients as intrinsically high redshift supernovae
will be facilitated by the deeper r and g-bands observations (24.7 and
25.0 magntiudes, respectively, for a single visit).

If supernovae from massive Population III stars are still forming at 
$z \sim 4$, as suggested by Fig.~\ref{fig:final_popIII_sfr}, these
sources will likely be well within LSST capabilities, thanks to a
further reduction in the distance modulus and the increased
sensitivity in the r band compared to the i band: These two factors
will yield an overall gain of more than 1 mag, and a $5 \sigma$
detection of a SN2006gy-like source at $z \sim 4$ will require
stacking of only 5 images in the g-band.

Our prediction for the \emph{observed} pair instability supernova rate
is plotted in Fig.~\ref{fig:pisn_rate} (for $v_{\rm wind}=60$ km~s$^{-1}$) 
and shows that about one supernova per year per $\sim 400$ deg$^2$ of sky
is expected at $z=6$, with the rate going down to one per $\sim 1250$
deg$^2$ at $z = 5$. The rate at $z \sim 5$ is high
enough to yield up to $\sim 100$ supernovae over the LSST
lifetime. Extrapolating the rate for g-dropouts ($z \sim 4$)
supernovae, we still expect a few events per year.

Establishing the high-redshift nature of these transients is expected
to be relatively simple, as no other transient objects are known to
possess such peak luminosities and light curve properties. A possible
source of contamination arises from pair instability supernovae
formed from the merger of two massive, metal-enriched stars
\citep{portegieszwart07}. Detailed estimates for the pair instability
supernova rate from this channel are still lacking.  Given the rarity of
2006gy-like supernovae in the local universe, the contamination is not
expected to be severe, below 1 event yr$^{-1}$.

The detection of Population III supernovae will be more difficult if
the progenitor stars are only moderately massive (typical masses $30~
\mathrm{M_{\sun}}$). {Still up to $6 \times 10^{51}$ ergs can by
  released by hypernovae with an initial mass of
  $20-25~\mathrm{M_{\sun}}$ \citep{nomoto08}.} However, the supernova
rate will be higher, potentially tens of supernovae per deg$^2$
assuming multiple stars per halo, and the peak luminosity of
hypernovae might be fainter than that of a pair instability explosion
\citep{nomoto03}. This will put them outside the reach of LSST, and a
specific deep survey should be planned to search for them. First, a
more detailed prediction for their light curve is needed. The
best-case scenario for supernovae detection would be if multiple, very
massive Population III stars formed per halo.

\section{Conclusion}\label{sec:conc}

In this paper, we investigated the possibility of a long tail in the
epoch of Population III star formation, extending to $z \leq 7$. This
epoch has been suggested to extend to redshifts after reionization
because of inhomogeneous IGM metal enrichment and a finite transition
from Pop~III to Pop~II modes of star formation
\citep[e.g.,][]{scannapieco03,jimenez06,tornatore07,shull08}. Our
simulations reach a mass resolution sufficient to identify Population
III star formation driven by H$_2$ cooling at $z \leq 26$ in halos
with $M \geq 10^6~ \mathrm{M_{\sun}}$. At the same time, they cover a
sufficiently large volume ($10^3~ \mathrm{Mpc^3}$) to be
representative of the Universe at $z \gtrsim 5$ for the halo mass
scale relevant to Population III formation (see \citealt{bagla05}).
Coupling the dark matter only dynamics with conservative assumptions
on self-enriched (genetic) and wind enriched scenarios, we show that
pockets of metal-free stars clearly exist well after the end of
reionization, although with a small number density (one new halo
Gpc$^{-3}$~yr$^{-1}$ formed at $z=6$). Our results confirm the
Population III star formation trend identified by \citet{tornatore07}
for $z \lesssim 6$, although their mass resolution was insufficient to
evaluate self-enrichment from minihalos, which might explain the
different behaviour at $z>6$ (see Appendix~\ref{sec:validation}).
However, our inferred Population III star formation rate assumes one
very massive star per halo and is smaller by a factor $\sim 100$ at $z
\sim 6$ compared to that plotted in Fig.~1 of \citet{tornatore07}. If
we were to assume instead that a fraction $\epsilon \sim 10^{-3}$ of
the metal-free gas in converted into Population III stars, then our
results would be much closer to those of \citet{tornatore07} at $z
\lesssim 6$.

While the $z \lesssim 10$ Population III star formation rate is
several orders of magnitude lower than that of metal-enriched stars,
the presence of metal-free gas at $z \lesssim 6$ opens an interesting
window for detecting Population III stars. If their expected initial
mass function is peaked toward very massive stars ($M \approx 100~
\mathrm{M_{\sun}}$) that could thus explode as hypernovae or
pair-instability supernovae, their expected rate (see
Fig.~\ref{fig:pisn_rate}) and peak luminosity of $m_{\rm AB} \lesssim
26$ at $z\lesssim 5$ suggest that future ground-based searches with
LSST can probe this mode of star formation. We predict one $z\approx
5$ Population III supernova per year per 1250 deg$^2$. We suggest that
large-area blind searches will be necessary to find these objects, as
the formation sites for Population III stars are distributed almost
uniformly on the sky, far ($\sim 150~h^{-1}$~kpc equivalent to $\sim
5$ arcsec) from the nearest (dwarf) galaxy. A prediction of our model
for metal-free supernovae is that no host galaxy will be present.
Furthermore, the two point correlation function of these supernovae
will show that these sources are significantly less biased than the
general population of $\sim 10^8 \mathrm{M_{\sun}}$ dark matter halos
at that redshift (see Fig.~\ref{fig:nearest_neighbors}).


All our assumptions about the enrichment of the IGM and the formation
rate of metal-free halos are robust. Thus, the presence of pristine gas
available for star formation at $z\lesssim 6$ is a very strong
prediction of our modeling. However, translating the amount of gas
available into an expected supernova rate carries a major uncertainty,
namely the Population III initial mass function and the star
formation efficiency. These are highly debated topic, in particular
the IMF of Population III stars. In fact some tension between
theory/simulations and observations is present \citep{tumlinson04,tumlinson06}. 
In addition, \citet{yoshida07} found with a 1-D model for cooling of
metal free gas in the presence of a strong UV background, that the IMF of
Population III stars is shifted toward smaller masses compared to
formation at higher redshift. Under such conditions, it is unclear
whether fragmentation will lead to multiple Population III
formation. Properly addressing the cooling process of these halos is
therefore a fundamental step toward evaluating the Population III
formation rate at the edge of the reionization epoch.  These issues 
will be examined in a follow-up paper.

\acknowledgements 

We thank \v{Z}eljko Ivezi\'{c} for {helpful} discussions {and
  an anonymous referee for useful suggestions}. MT and JMS acknowledge
support from the University of Colorado Astrophysical Theory Program
through grants from NASA (NNX07AG77G) and NSF (AST07-07474). MS
acknowledges support from NASA (NAG5-12458). Numerical simulations were
carried out on the Royal Cluster at STScI.

\bibliography{popIII}{}

\clearpage

\appendix
\section{Validation and Convergence}\label{sec:validation}
 
To correctly evaluate the number density of metal-free halos around
the epoch of reionization, it is necessary to ensure that the numerical
simulations we used have a sufficient resolution. Strong winds can
originate only from relatively massive halos, and are thus relatively
easy to identify. However, self-enrichment is more difficult to model
because it may arise primarily from minihalo progenitors. Identifying
these halos requires a high mass resolution (single particle mass below 
$\sim 5 \times 10^{4}~\mathrm{M_{\sun}}$).

\begin{figure} 
  \plotone{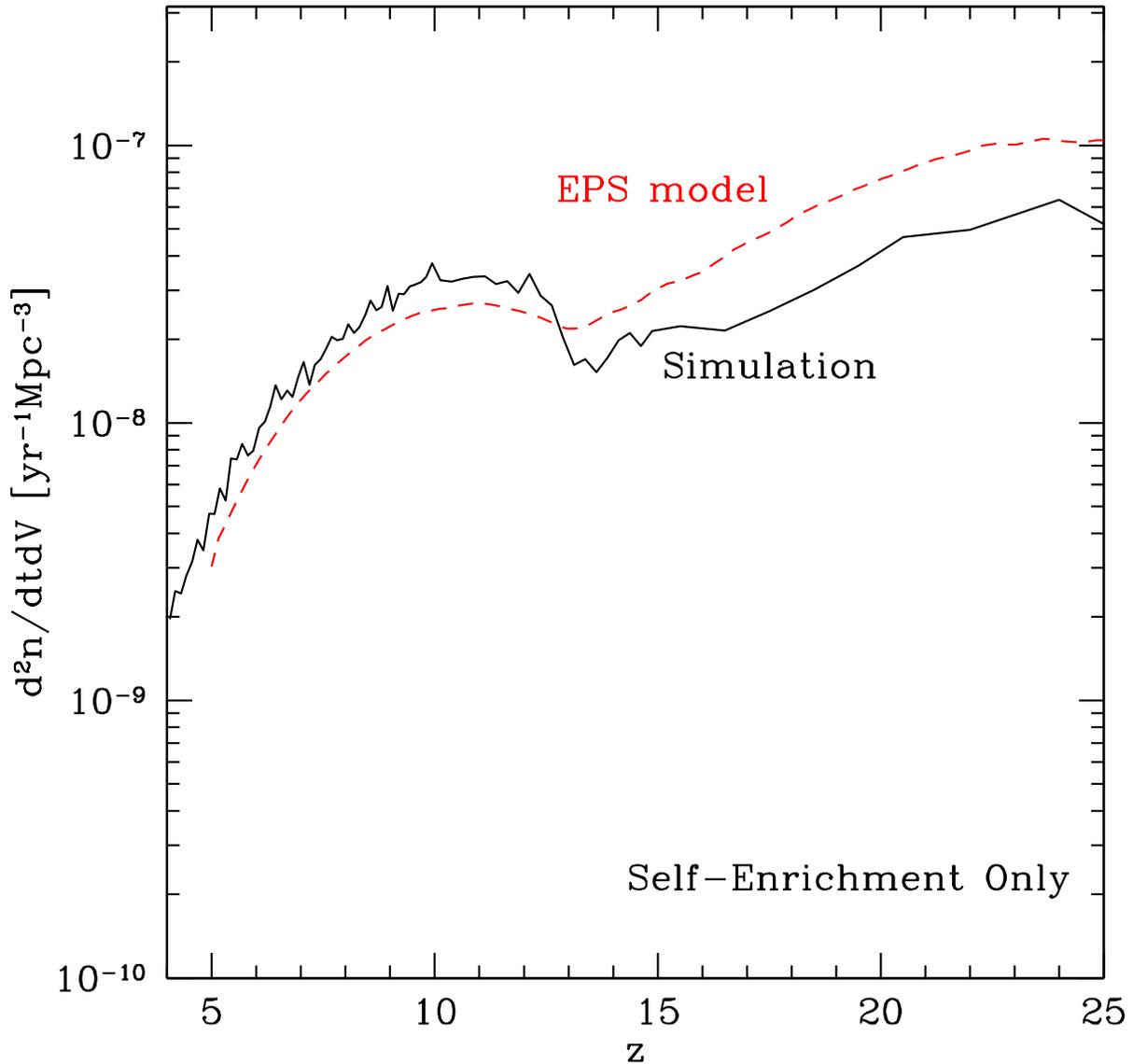}\caption{Solid black line: formation rate of halos
  that have not experienced self-enrichment (number of new halos
  formed per proper unit time (yr) per unit volume (comoving
  Mpc$^3$)). The red dashed line is the prediction for the same
  quantity based on the Extended Press-Schechter (``EPS'') model of
  \citet{ts09}. There is a remarkable agreement especially at redshift
  $z \lesssim 10$. Winds are not taken into account in this figure.}
\label{fig:genetic_pristine}
\end{figure}

\begin{figure} 
  \plotone{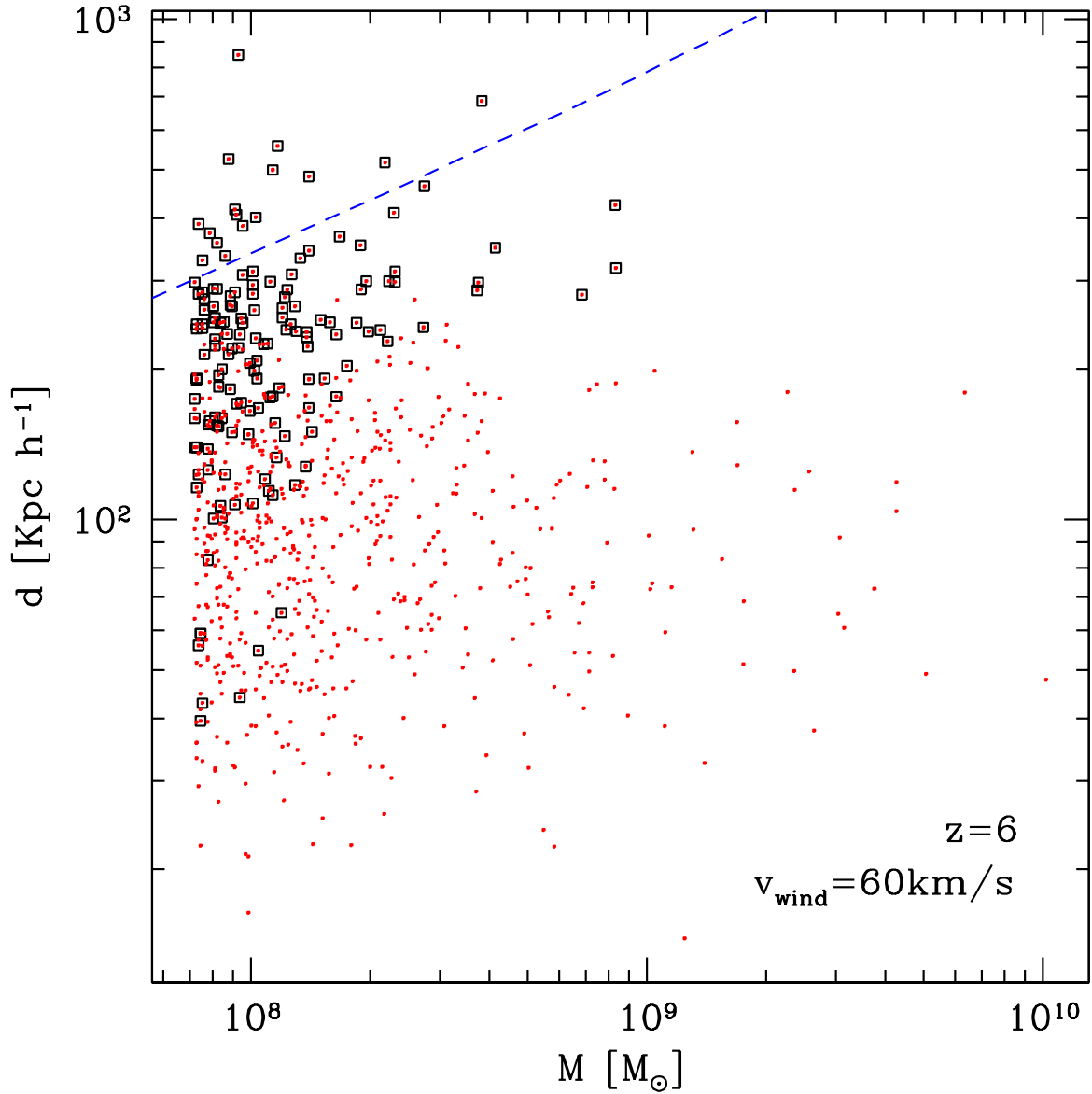}\caption{Comoving distances and masses of the
    nearest-neighbor halo of non self-enriched (red dots) and truly
    pristine halos (black squares) at redshift $z=6$ for a wind speed
    $v_{\rm wind}=60$ km~s$^{-1}$. A significant fraction of halos that
    were not self-enriched have neighbors too distant to pollute them
    with their metal winds. The blue dashed line gives the average
    distance of halos of mass $M$ under the assumption of a uniform
    distribution, highlighting the effect of clustering. Truly
    pristine halos are significantly less clustered than their metal
    enriched counterparts.}\label{fig:nearest_neighbors}
\end{figure}

\begin{figure} 
  \plotone{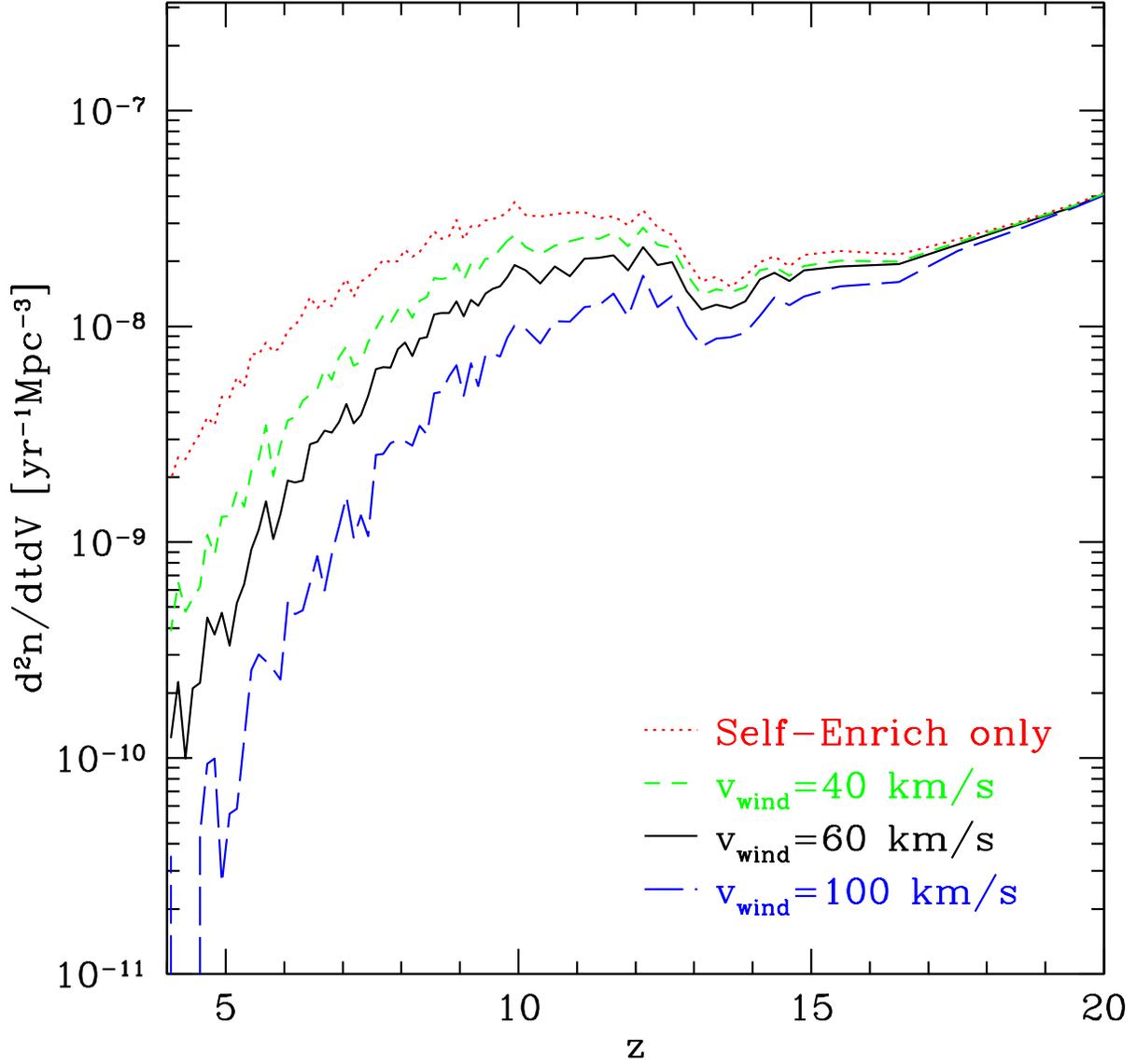}\caption{Population III halo formation rate after
  accounting for wind pollution ({short dashed green line $v_{\rm wind}=40$  km~s$^{-1}$; solid black line $v_{\rm wind}=60$ km~s$^{-1}$; 
  long dashed blue line $v_{\rm wind}=100$ km~s$^{-1}$}) compared
  to the formation rate of halos that were not self-enriched (dotted
  red line). The red dotted line is also equivalent to assuming
  $v_{\rm wind}=0$.}\label{fig:final_popIII_sfr}
\end{figure}

\begin{figure} 
  \plotone{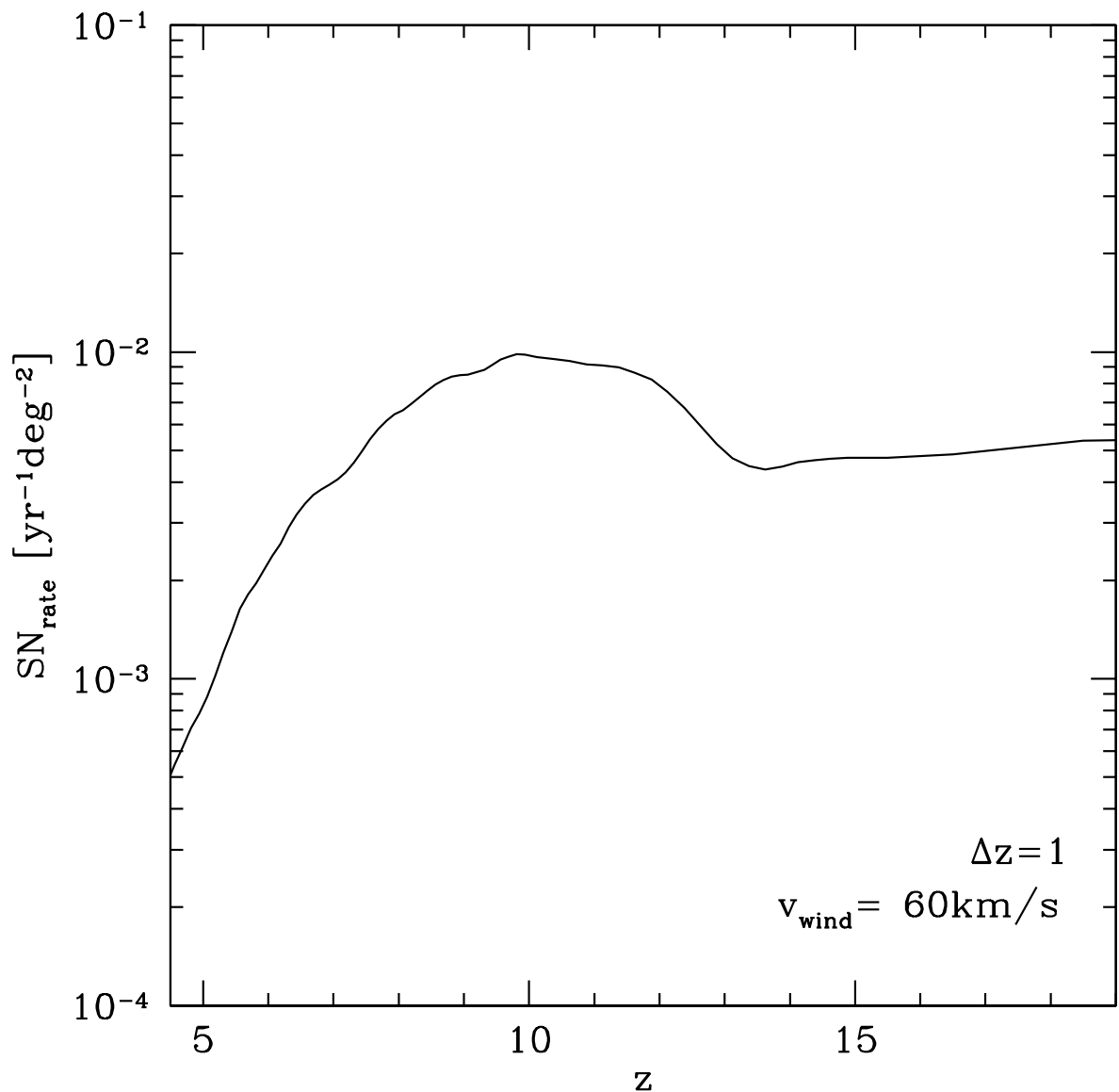}\caption{Observed Population III supernova rate
  per year per deg$^2$ as a function of redshift for our standard
  model of wind propagation ($v_{\rm wind}=60$ km~s$^{-1}$). This has been
  obtained assuming one Population III supernova per halo. The field
  of view depth is $\Delta z = 1$ centered at redshift $z$, a
  choice representative for typical Lyman-Break dropout
  searches.}\label{fig:pisn_rate}
\end{figure}

\begin{figure} 
  \plotone{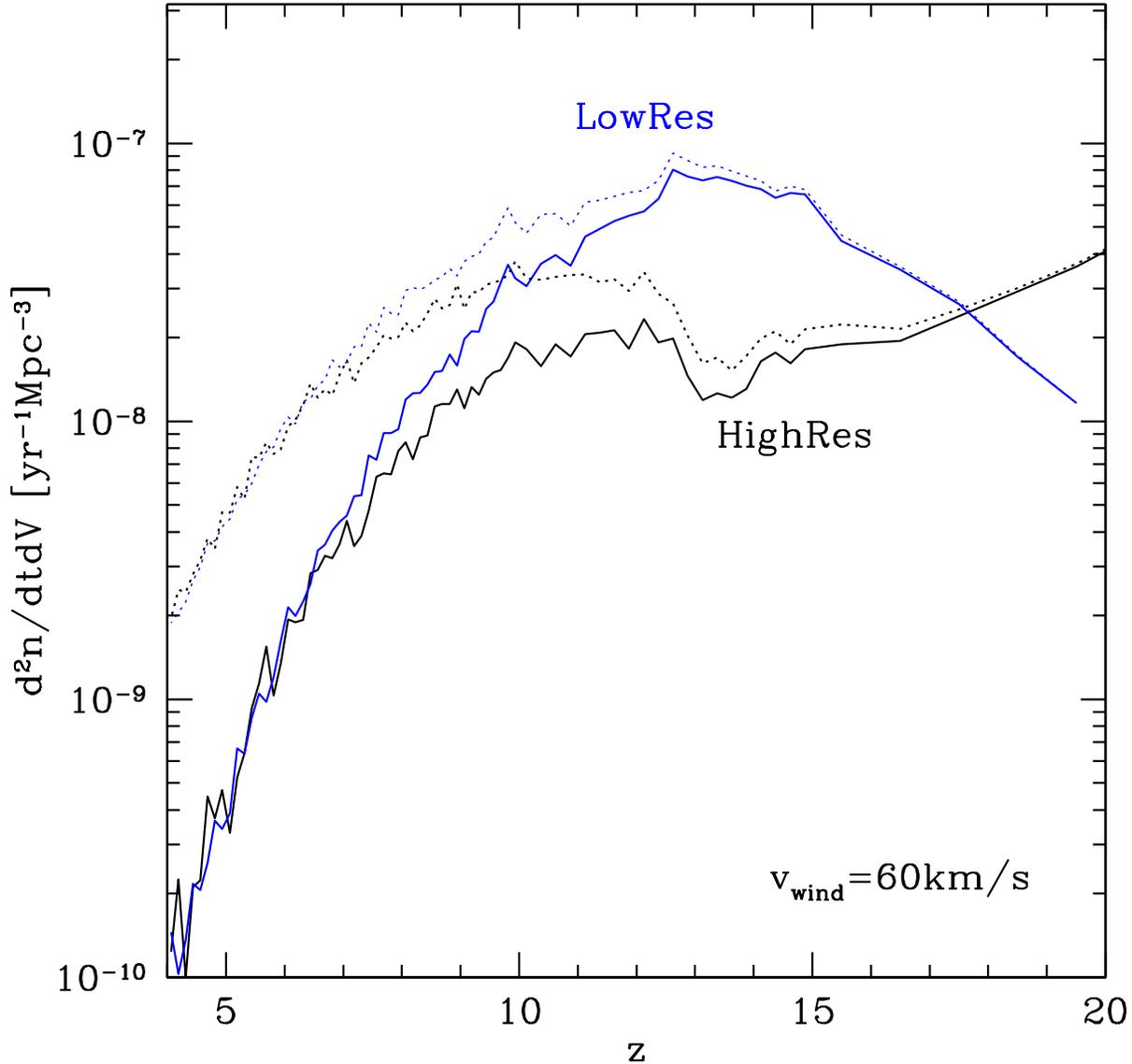}\caption{Population III halo formation rate
  obtained with our standard wind prescription ($v_{\rm wind}=60$ km~s$^{-1}$)
  from two cosmological simulation, a ``Low-Resolution'' $N=512^3$ run
  (solid blue line) and a ``High-Resolution'' $N=1024^3$ run (solid
  black line). The dotted blue and black lines represent the rate in
  presence of self-enrichment only. The results of the two runs agree
  well for $z \lesssim 8$. At higher redshift ($8 \lesssim z \lesssim
  17$), the low resolution run is not able to follow self-enrichment
  and thus over-estimates the formation rate of metal-free
  halos.}\label{fig:validation}
\end{figure}

A strong hint that our numerical results have reached convergence with
respect to self-enrichment comes from the excellent agreement of the
Population III halo formation rate with the analytic model for
self-enrichment presented in \citet{ts09} (see
Fig.~\ref{fig:genetic_pristine}). In addition, we compare here our
production run with $N=1024^3$ and a box size of $7~h^{-1}$~Mpc 
(single particle mass of $3.4 \times 10^{4}~ \mathrm{M_{\sun}}$) to 
the results of a control run at lower resolution, with $N=512^3$ and 
box size of $10~h^{-1}$~Mpc (single particle mass of 
$7.9 \times 10^{5}~ \mathrm{M_{\sun}}$). In
Fig.~\ref{fig:validation} we plot the corresponding halo formation
rate for our reference model with a wind speed of $60$ km~s$^{-1}$. We
clearly see that at $z \lesssim 7$ the two simulations yield the same 
results. The relative difference in $d^2n/dt\,dV$ is below the variance
of the halo number counts in a single snapshot. Note that at higher
redshift ($z\sim 15$) the lower resolution run exhibits a peak in
Population III formation rate, which is a numerical artifact induced by
the inability to follow self-enrichment due to minihalos formed at $z
\gtrsim 20$. The resolution of this ``low resolution'' control run is
equal to the highest resolution run of \citet{tornatore07} suggesting
that their results might not be reliable at $z \gtrsim 10$.

\clearpage

\clearpage

\begin{table} \begin{center} \caption{Sensitivity of current and
      future facilities for Population III detection \label{tab:obs}}
\begin{tabular}{lccccr}
\tableline\tableline
Telescope/Instrument & Band & Sensitivity  & Integration Time & Field of view  & Time/($10^3$ deg$^2$) \\
\small{LSST} & i & $m_{AB}=26$ & 1200 s & 9.6 deg$^2$ & 35 h\\ 
\small{SUBARU/Suprime-Cam} & i &  $m_{AB}=26$ & 470 s& 0.25 deg$^2$ & 522 h\\
\small{HST/ACS} & F775W & $m_{AB}=26$& 800s &  11.3 arcmin$^2$& $7 \times 10^4$ h \\
\small{JWST/NIRCAM} & F090W & $m_{AB}=26$ & $<$ 30 s &9.7 arcmin$^2$ & 3093 h \\
\small{TMT/IRIS$^{a}$} & i & $m_{AB} = 26$& $\sim$ 30 s& 0.06 arcmin$^2$ & $5 \times 10^5$ h \\ 
\tableline 
\small{JWST/NIRCAM} & F090W & $m_{AB}=31$ & 65 h &9.7 arcmin$^2$ & N/A \\ 
\small{TMT/IRIS$^{a}$} & z & $m_{AB}=31$ & 34 h &0.06 arcmin$^2$ & N/A \\ 
\tableline
\end{tabular}
\tablecomments{Summary of the sensitivity for $4.5 \lesssim z \lesssim
  7$ Population III SNe searches using the current and future
  facilities listed in the first column. The second column reports the
  filter used, the third the $5 \sigma$ sensitivity reached in the
  time given in the fourth column. The fifth column gives the field of
  view and the last column the time needed to survey $10^3$ deg$^2$:
  SNe detection is maximally efficient with LSST given its superior
  field of view. Last two lines: a deep field with JWST NIRCAM or with
  a 30 m class telescope will approach the depth necessary for direct
  imaging of a cluster of Population III stars, if these exist.
  \emph{a}: TMT sensitivity is an estimate based on the ``TMT
  Instrumentation and Performance Handbook''
  (http://www.physics.uci.edu/TMT-Workshop/TMT-Handbook.pdf), as no
  exposure time calculator is currently available.}
\end{center}
\end{table}


\end{document}